\documentclass[journal]{IEEEtran}

\usepackage{cite}
\usepackage[pdftex]{graphicx}
\usepackage{xcolor, tikz}
\usepackage{amsmath, amsfonts, bm, scalerel}
\usepackage[ruled, vlined, commentsnumbered, linesnumbered]{algorithm2e}
\usepackage{algorithmic}
\usepackage{array, multirow}
\usepackage{url, hyperref}

\newcommand{\bx}{{\bf x}}
\newcommand{\etal}{{\em et al.}}       
\newcommand{\eg}{{\em e.g.}}           
\newcommand{\ie}{{\em i.e.}}           
\newcommand{\etc}{{\em etc.}}         

\begin{document}
\title{A Novel RL-assisted Deep Learning Framework\\for Task-informative Signals Selection and Classification for Spontaneous BCIs}

\author{Wonjun~Ko,~Eunjin~Jeon,~and~Heung-Il~Suk,~\IEEEmembership{Member,~IEEE}
\thanks{W. Ko was with the Department of Brain and Cognitive Engineering, Korea University, Seoul 02841 Korea Republic.}
\thanks{E. Jeon was with the Department of Brain and Cognitive Engineering, Korea University, Seoul 02841 Korea Republic.}
\thanks{H.-I. Suk was with the Department of Artificial Intelligence and the Department of Brain and Cognitive Engineering, Korea University, Seoul 02841 Korea Republic.}
}

\markboth{UNDER REVIEW}%
{Ko, Jeon, and Suk: A Novel RL-assisted DL Framework for Task-informative Signals Selection and Classification for Spontaneous BCIs}

\maketitle

\begin{abstract}
In this work, we formulate the problem of estimating and selecting task-relevant temporal signal segments from a single EEG trial in the form of a Markov decision process and propose a novel reinforcement-learning mechanism that can be combined with the existing deep-learning based BCI methods. To be specific, we devise an actor-critic network such that an agent can determine which timepoints need to be used (informative) or discarded (uninformative) in composing the intention-related features in a given trial, and thus enhancing the intention identification performance. To validate the effectiveness of our proposed method, we conducted experiments with a publicly available big MI dataset and applied our novel mechanism to various recent deep-learning architectures designed for MI classification. Based on the exhaustive experiments, we observed that our proposed method helped achieve statistically significant improvements in performance.
\end{abstract}


\IEEEpeerreviewmaketitle

\section{Introduction}
\IEEEPARstart{B}{rain}--computer interface (BCI) is an emerging technology that allows communicable pathways between a brain and an external device, \eg, a robotic arm, by measuring and identifying intention-reflected brain activities \cite{bhatti2019soft}. Generally, non-invasive BCI systems, commonly using electroencephalogram (EEG), are categorized into two types, \emph{evoked} and \emph{spontaneous} BCIs. While evoked BCIs exploit evoked potentials like P300, mostly induced by an external stimulus, spontatneous BCIs focus on internal cognitive processes such as event-related (de)synchronization (ERD/ERS). In this work, we focus on \emph{motor imagery} (MI) induced brain signals \cite{pfurtscheller2001motor}.

Since MI-EEGs are voluntarily inducible, MI-based BCIs show great values in the clinical and applicational standpoints. However, because of the self-inducing property and difficulty in consistently inducing spontaneous EEG signals for a period of time, the MI-EEG signals are highly likely to have not only MI-relevant information, but also irrelevant information in trials \cite{li2017relevant}, which is regarded as \emph{unreliable} EEG segments in the following description. Generally, in an MI-EEG acquisition protocol, self-induced MI-EEG data is obtained by presenting a cue-signal (\eg, left-arrow sign to imagine the movement of left-hand, right-arrow sign to imagine the movement of right-hand, \etc) \cite{lee2019eeg}. Therefore, the acquired EEG data could have unreliable segments, when the subject does not fully concentrate during the MI-EEG acquisition because of lack of familiarity in BCIs or uncomfortable condition, \eg, long-calibration time. Further, the MI-EEG can also have different physiological noise, \eg, heartbeat, eyeball movement, \etc~\cite{li2017relevant}. Thus, it is not reasonable to have a complete \emph{reliability} to the acquired EEG trials. 

\begin{figure}[t]\centering
	\begin{minipage}[t]{.23\textwidth}
		\begin{center}
			\includegraphics[width=1\linewidth]{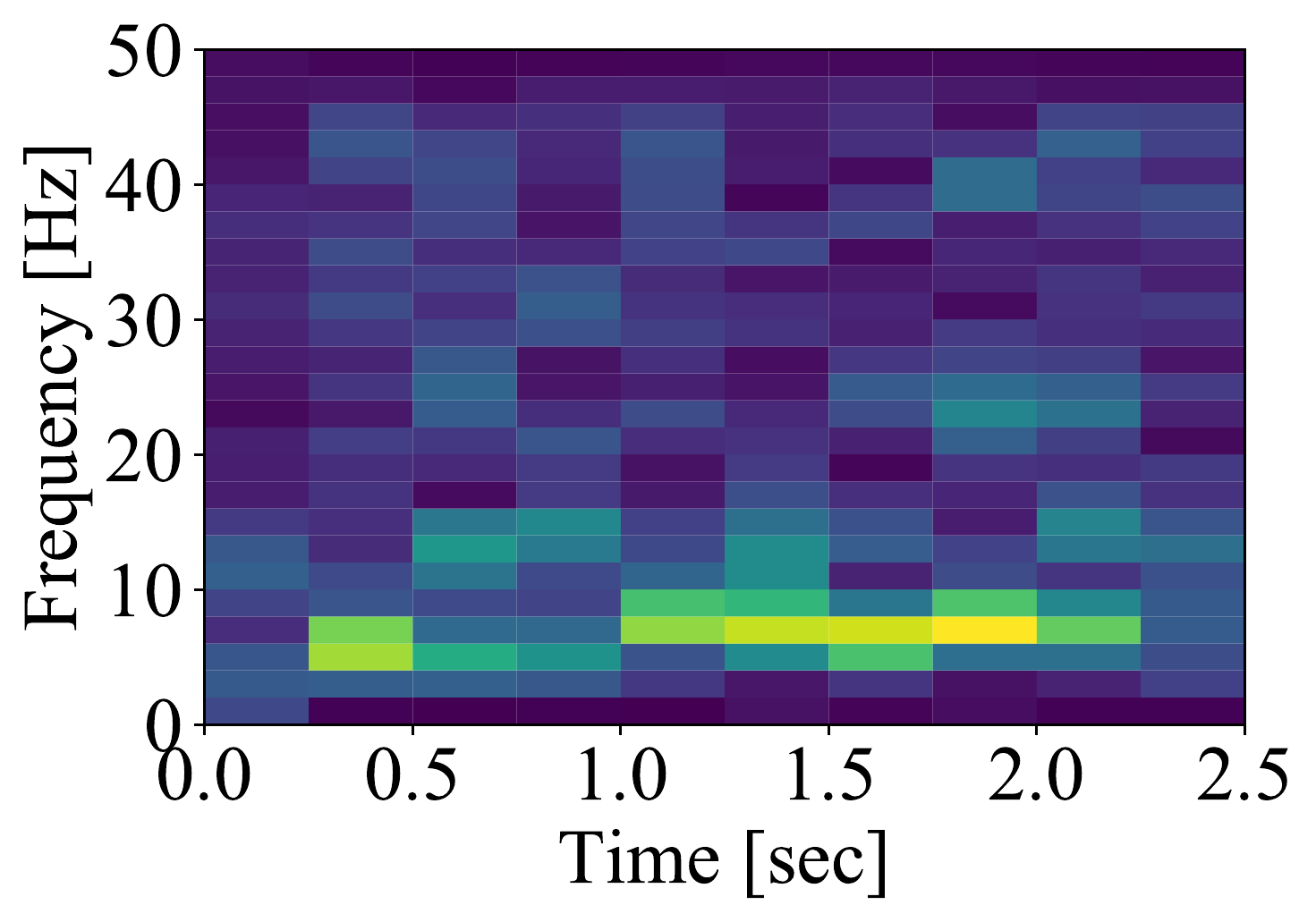}
		\end{center}\vspace{-5pt}
		\footnotesize
		\centering{(a) Subject \#28}
	\end{minipage}\hspace{10pt}
	\begin{minipage}[t]{.23\textwidth}
		\begin{center}
			\includegraphics[width=1\linewidth]{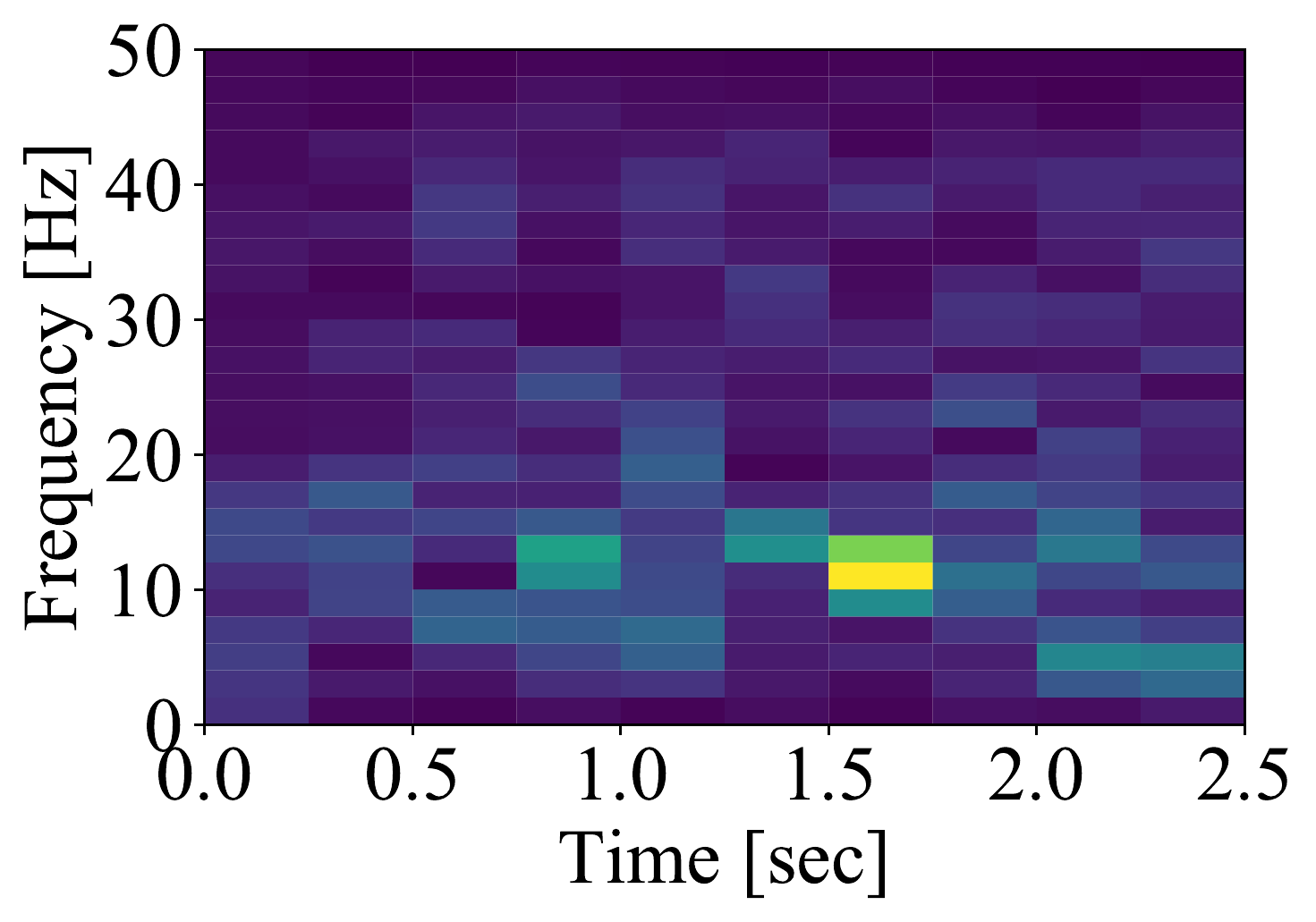}
		\end{center}\vspace{-5pt}
		\footnotesize
		\centering{(b) Subject \#11}
	\end{minipage}
	\caption{Comparison of the power spectrogram (by short-time Fourier transform) of the C4 channel in left-hand motor imagery trials from two subjects. There is a clear and lasting pattern in the range of $\mu$-band for the Subject \#28 (a). However, there is no evident and lasting activation pattern in neither of $\mu$- nor $\beta$-bands for the Subject \#11 (b).}
	\label{fig: stft}
\end{figure}

As an example, Fig. \ref{fig: stft} compares the power spectrogram of the C4 channel in left-hand MI trials from two subjects. Many neurophysiological studies on physical or imagery movements \cite{mcfarland2000mu, lawhern2018eegnet} have consistently witnessed that MI-caused signal patterns are observed in $\mu$ (8-12Hz) and/or $\beta$ (12-30Hz) bands, even though there are not generic frequency ranges that can be applicable to all subjects and there are high variations in signal patterns among subjects and even among sessions of the same subject. In the spectrogram of the Subject \#28, the high power pattern in the near $\mu$-band is observed and lasting for a period of time. However, such evident patterns are not observable in the spectrogram of the Subject \#11. Thus, the typical machine-learning algorithms including recent deep learning methods \cite{schirrmeister2017deep, lawhern2018eegnet, ko2020multi, kwon2019subject, blankertz2008optimizing, ang2008filter} that exploit the whole signals of trials for model training and intention identification may not be equally applicable to those subjects.

\begin{figure}[t]\centering
	\includegraphics[width=.65\linewidth]{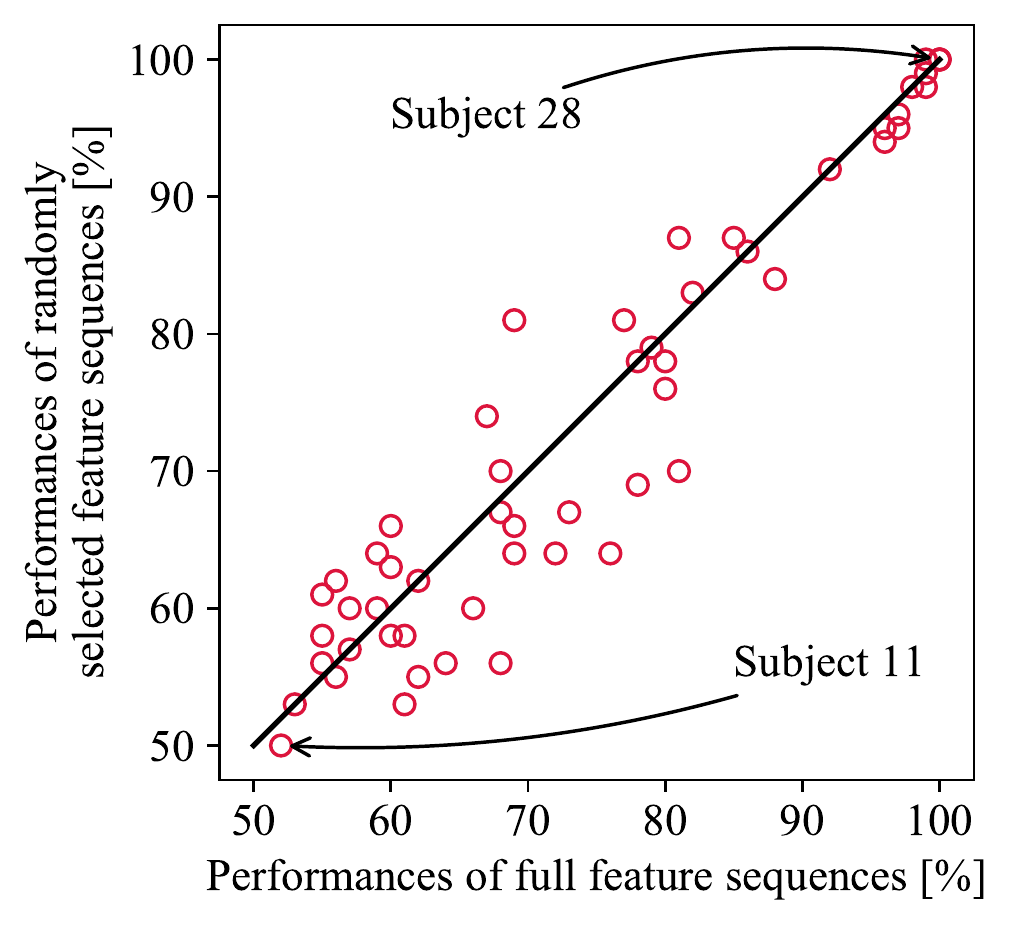}
	\caption{Performances comparison of a predictive model by training with either the full signals (mean accuracy: 74.39$\pm$15.59\%) and with the randomly selected signal segments (mean accuracy: 73.43$\pm$15.92\%).}
	\label{fig: compare}
\end{figure}

There have been recent studies that considered the unreliability of information in features or raw data in training predictive models \cite{li2017relevant, zhang2019feature, liu2017quality}. Among them, Li \etal~\cite{li2017relevant} suggested that training predictive models with the full EEG signals of BCI trials is not necessarily helpful to enhance classification performance in MI-BCIs. Inspired by their work, we performed a preliminary study to compare the performance changes between two models trained and tested with (1) full signals (FM) and (2) randomly masked out signals in time, thus discarding the respective features (RM) for individual subjects. The resulting plot is shown in Fig. \ref{fig: compare}. Interestingly, we could observe that for many subjects, their respective performance with RM was almost the same with or higher than that with FM.
Based on that result, we hypothesize that rather than extracting features from the full signals in a trial, it would be effective to select intention-related signal segments, \ie, to discard intention-unrelated or noisy signals, and to use them only for feature representation and the ensuing classifier learning. 


In the meantime, while MI-EEGs are obtained by the general protocol, there is no way to know whether the given temporal signal is MI-relevant or not. In other words, we cannot have any information about MI-relevancy for acquired EEG signals explicitly. Thus, we formulate the problem of selecting MI-relevant signal segments without any \emph{supervision} in the form of a Markov decision process and tackle it via \emph{reinforcement learning} (RL) \cite{mnih2016asynchronous} systematically. To our best knowledge, this is the first work that proposes RL-based intention-related signal segments selection and jointly learning feature representation and a classifier in a unified framework. 

The main contributions of our work are as follows:
\begin{itemize}
	\item First, we tackle the problem of estimating and selecting \emph{reliable} signals in MI-EEG, which can be an important issue to practical usage of BCI, by formulating in an RL framework.
	\item Second, we devise an actor-critic model for MI-based BCI and define a novel reward function.
	\item As our proposed of the RL-based feature vectors selection over time is modular, it is easy to plug into the existing deep-learning architectures with minor modification, and thus to help enhance classification performance.
	\item In our experiments over a big MI dataset, we achieved statistically significant performance improvements with our proposed method injected in various deep networks, further outperforming other comparative methods in the literature.
\end{itemize}

This paper is organized as follows: Section \ref{sec: related work} reviews the previous studies on EEG decoding methods including deep learning approaches and MI-relevant EEG trials selection.
In Section \ref{sec: methods}, we propose an MI-relevant EEG signal segments selection method in an actor-critic framework \cite{mnih2016asynchronous} and describe our objective optimization strategy with a novel reward function. Section \ref{sec: experiments} describes the EEG dataset, experimental settings, and quantitative results by comparing with the existing methods in the literature. We then analyze the results to further validate the effectiveness of our method in Section \ref{sec: analysis}, and finally summarize our work in Section \ref{sec: conclusion}.

\section{Related Work}
\label{sec: related work}
Over the past decades, a common spatial pattern (CSP) algorithm \cite{blankertz2008optimizing} and its variants \cite{ang2008filter, suk2012novel} have been studied most actively for MI-EEG decoding by focusing on spatial filters learning such that the signals are transformed and dimension-reduced to be better discriminative. In particular, Ang \etal~\cite{ang2008filter} band-pass filtered MI signals before applying CSP, thereby representing spatio-spectral features of EEG signals. Suk and Lee \cite{suk2012novel} proposed a Bayesian framework to jointly optimize the spectral filters and spatial filters in a unified framework by defining frequency bands as random variables.

Meanwhile, deep learning methods have achieved promising results in EEG signal decoding studies \cite{zhang2019survey, gu2020eeg}. For instance, Schirrmeister \etal~\cite{schirrmeister2017deep} proposed various convolutional neural networks (CNNs) for MI classification, \eg, Shallow ConvNet and Deep ConvNet. Ko \etal~\cite{ko2018deep} proposed an interesting recurrent spatio-temporal CNN architecture. Lawhern \etal~\cite{lawhern2018eegnet} proposed an EEGNet that exploited depth-wise convolutional layers and separable convolutional layers \cite{chollet2017xception} for reducing tunable parameters, thereby learnable with a limited number of EEG samples. Zhang \etal\ \cite{zhang2018cascade} proposed Parallel CRN and Cascade CRN, combined recurrent neural network (RNN) and CNN to extract spatio-spectral features of MI-EEG. Further, Kwon \etal\ \cite{kwon2019subject} also proposed multi spectral-spatial feature representation (SSFR) using spectral filtering and CNN for MI decoding on both subject-dependent and independent manners. More recently, Ko \etal~\cite{ko2020multi} devised multi-scale neural network (MSNN), which learns multi-scale (in frequency) feature representations of EEG signals, and presented its applicability to various EEG-based applications.

Unlike most of the existing methods that focused on spatial or spatio-spectral feature extraction with no attempt to find task-relevant EEG trials or signals in trials, Fruitet \etal~\cite{fruitet2012bandit} focused on task-related trials selection by formulating it as a multi-armed bandit problem \cite{sutton2018reinforcement}. In particular, given an EEG trial, their method estimates the confidence of containing task-relevant information compared to idle state EEG signals. Recently, Li \etal~\cite{li2017relevant} proposed spectral component CSP (SCCSP) to select MI relevant EEG trials. Specifically, they conducted independent component analysis \cite{hyvarinen2000independent} on bandpass-filtered signals for MI-relevant and MI-irrelevant components extraction on each class independently. The extracted components were then used for MI-relevant EEG trials selection from the training dataset, based on which they ran CSP for feature extraction and trained a classifier.

\begin{figure*}[t]\centering
	\begin{minipage}[t]{.65\textwidth}
		\begin{center}
			\includegraphics[width=1\linewidth]{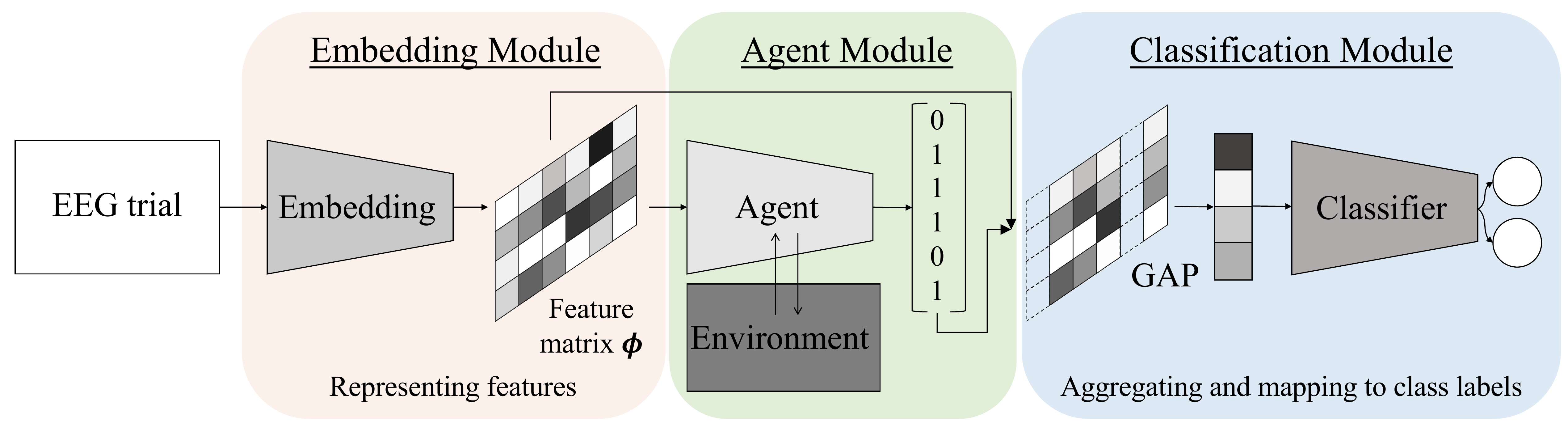}
		\end{center}
		\footnotesize
		(a) Illustration of our proposed framework, which is composed of three basic modules, namely, an input signals embedding module, an agent module for signal segments selection, and a classification module. The agent interacts with an environment in the process depicted in Fig. \ref{fig: framework}(b). 
	\end{minipage}
	\hspace{5pt}
	\begin{minipage}[t]{.32\textwidth}
		\begin{center}
			\includegraphics[width=1\linewidth]{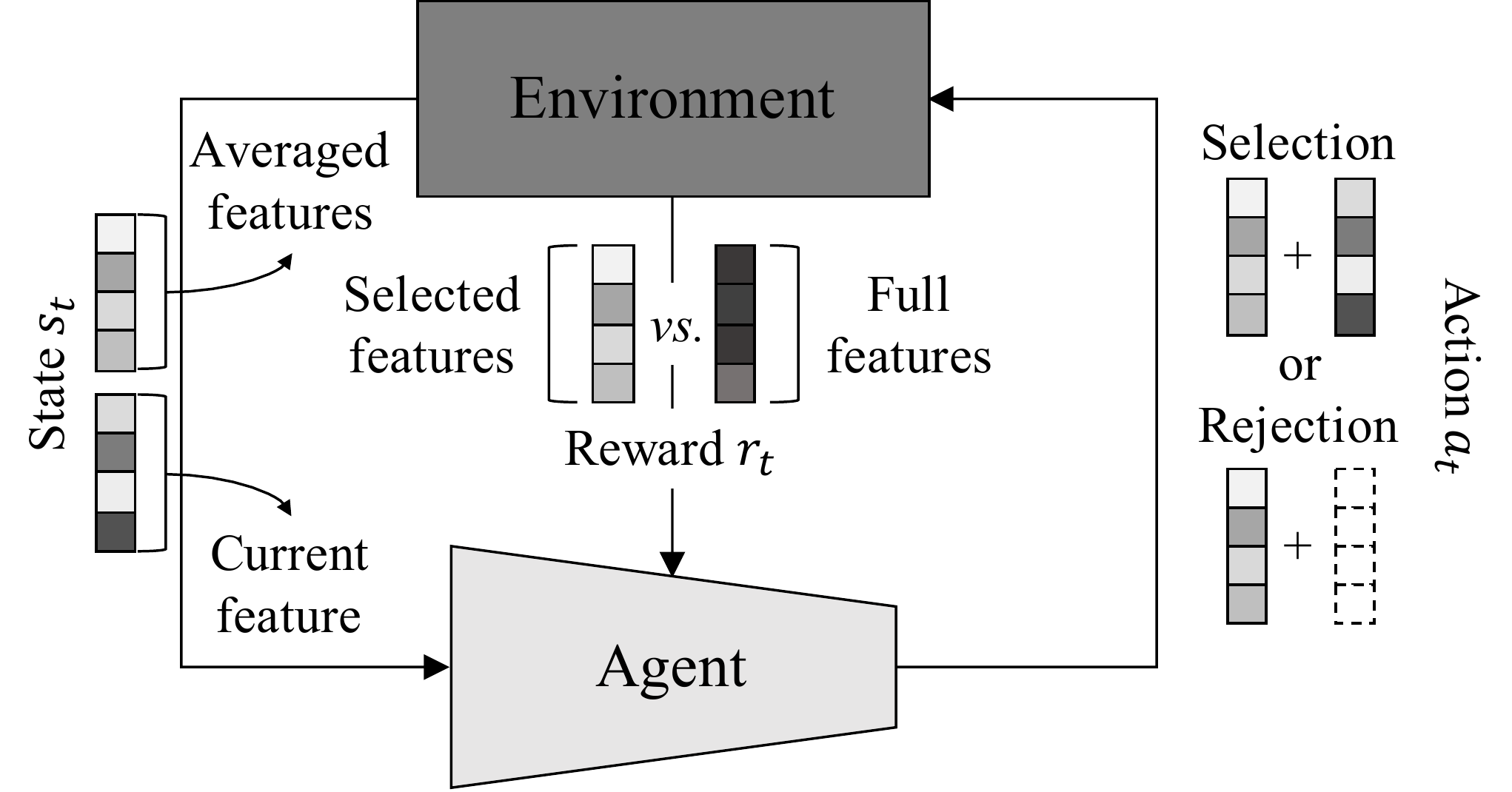}
		\end{center}
		\footnotesize
		(b) A schematic representation of the interaction between the agent and the environment for the task of signal segments selection. For the definitions and details about state $s_{t}$, action $a_{t}$, and reward $r_{t}$, refer to the contexts in the main body.
	\end{minipage}
	
	\caption{A graphical overview of the proposed framework and the internal mechanism of the proposed RL module for task-related informative feature vectors selection over time}.
	\label{fig: framework}
\end{figure*}

Our method can be comparable to their methods in the sense of concerning MI-relevant signals selection in a framework. First, we consider signal segments selection in each trial, rather than selecting trials in a dataset. That is, we can still use the whole trials in a training set by allowing to maximally utilize all the available samples. Second, when comparing with Fruitet \etal's work~\cite{fruitet2012bandit}, our method do not require idle state EEG trials, which otherwise could be great limitation as requiring additional time for data acquisition, thus causing a longer calibration time accordingly. Further, unlike Li \etal's work \cite{li2017relevant} of learning baseline components, which are used to determine MI-relevance of EEG signals, feature extraction and classifier learning separately, we devise a systematically integrated framework for feature representations learning, estimation and selection of MI-relevant feature vectors of signal segments, and  classifier learning in a unified framework. It is also noteworthy that those modules are jointly optimized in an end-to-end manner. Throughout the paper, we use the terms of signal segments and temporal feature vectors of EEG signals interchangeably.

\section{Methods}
\label{sec: methods}
In this section, we define the MI-relevant EEG signal segments selection problem, and formulate it in a novel framework where a reinforcement-learning induced module plays a vital role for performance enhancement. The proposed framework has three main modules as schematized in Fig. \ref{fig: framework}(a). Given a sequence of signals in a trial $\bx=\left\{x_{1},\dots,x_{T}\right\}\in\mathbb{R}^{C\times T}$, where $C$ and $T$ denote, respectively, the number of channels and timepoints, it first passes through an embedding network for features representation. The represented feature vectors are then fed into our novel agent module to estimate their task-relevancy and to select the informative signal segments for the target task. Finally, a classifier makes a decision for the task, \ie, MI classification, using the selected feature vectors over time.



\subsection{Embedding Network}
Notably, this module is flexible with many kinds of network architectures, varying from the existing ones in the literature to newly customized networks. In our experiments, we exploit the existing CNN architectures, namely, ShallowNet \cite{schirrmeister2017deep}, DeepNet \cite{schirrmeister2017deep}, EEGNet \cite{lawhern2018eegnet}, and MSNN \cite{ko2018deep}. Basically, these architectures were proposed by different research groups and presented their superiority or validity in their respective experiments over various datasets. In the following, we denote an embedding network for feature representation as $\phi(\cdot; \theta_{\phi})$ with a tunable parameters $\theta_{\phi}$.

\subsection{Agent Network}
We introduce a learnable agent that adaptively and automatically selects task-relevant feature vectors of EEG signals over time in a trial without \emph{supervision}, as there is no explicit way of observing such information in a trial. For the feature vectors $\boldsymbol{\phi}=\left\{\phi_{1},\dots,\phi_{T'}\right\}\in\mathbb{R}^{D\times T'}$ of the input signals, where $D$ is the dimension of feature vectors, we devise a method for automatic selection of signal segments over time $t\in\mathcal{S}$, $\mathcal{S}\subset\left\{1,\dots,T'\right\}$, such that the selected feature vectors $\left\{\phi_{t}\right\}_{t\in\mathcal{S}}$ carry the most information related to the user's intention, induced by means of MI. However, as MI involves an internal cognitive process in a brain, and thus there are no clear labels, \ie, informative or non-informative, for signals at which timepoints they actually include the intention-related information.

Here, we formulate the problem of informative feature vectors selection of signals in a Markov decision process \cite{sutton2018reinforcement} and devise an RL-assisted module to enhance the MI-EEG classification performance. Specifically, an agent interacts with the environment defined with a given MI-EEG trial via a sequence of states (defined with the set of feature vectors represented by an embedding network $\phi$), actions (selection or rejection), and rewards (effects of making specific actions, \ie, decisions) over time, as illustrated in Fig. \ref{fig: framework}(b).

In order to demystify our method, we define states, actions, and rewards as follows: 
\subsubsection{State}
A state $s_t$ $(t=1,\dots,T')$ in our work is represented as a continuous vector constructed by concatenating the aggregated feature vectors of the selected up to the previous time point, \ie, $\mathrm{AGG}\left(\left\{\phi_{i}\right\}_{i\in\mathcal{S}_{t-1}}\right)$ and the same one but further including the feature vector of the current time $t$, \ie, $\phi_{t}$ as follows:
\begin{equation}
	s_{t} = \mathrm{Concat}\left(
	\begin{array}{c}
		\mathrm{AGG}\left(\left\{\phi_{i}\right\}_{i\in\mathcal{S}_{t-1}}\right), \\
		\mathrm{AGG}\left(\left\{\phi_{i}\right\}_{i\in\mathcal{S}_{t-1}} \cup \{\phi_{t}\}\right)
	\end{array}
	\right)
\label{eq: state}
\end{equation}
where $\mathcal{S}_{t-1}$ is an index set of the selected feature vectors up to the time $t-1$. the operators of $\mathrm{Concat}$ and $\mathrm{AGG}$ denote, respectively, a vector concatenation operator and an aggregation operator. In our work, we use a mean aggregator defined as 
\begin{equation}
\mathrm{AGG}\left(\left\{\phi_{i}\right\}_{i\in\mathcal{S}_{t}}\right) = \frac{1}{\left|\mathcal{S}_{t}\right|}\sum_{i\in \mathcal{S}_{t}}\phi_{i}
\end{equation}
where $|\mathcal{S}_{t}|$ is a cardinality of the set $\mathcal{S}_{t}$.



\subsubsection{Action}
An action space $\mathcal{A}$ is defined to make it possible for the agent to select (1) or reject (0) the sequence of feature vectors over time and we are interested in finding an optimal action sequence to maximize the expected rewards. 
Concretely, referring to the current state $s_{t}$ that involves the comparative information of both aggregating and non-aggregating the feature vector of the current time $t$ with the features of the earlier selected, it estimates the effect of the current feature vector to increase the resulting expected rewards. Based on the agent's action, the set $\mathcal{S}_t$ is updated as follows: 
\begin{equation}
	\mathcal{S}_t = \left\{
		\begin{array}{ll}
			\mathcal{S}_{t-1}\cup \{t\} & \text{if } a_{t}=1 \text{ (selection)} \\
			\mathcal{S}_{t-1} & \text{otherwise } \text{ (rejection)}
		\end{array}
	\right.
	\label{eq:action}
\end{equation}

\subsubsection{Reward}
In order to define the rewards with respect to actions made by the agent, we first define the base information by taking a global average pooling (GAP) \cite{lin2013network} over the whole feature vectors over time in a trial as follows:
\begin{equation}
\mathbf{f}_\mathrm{GAP}=\mathrm{AGG}\left(\left\{\phi_{t}\right\}_{t\in\{1,\dots,T'\}}\right)
\label{eq: gap}
\end{equation}
and calculate the classification loss $\mathcal{L}_\mathrm{GAP}$ as a criterion.
Then, the reward $r_{t}$ with respect to the current action $a_{t}$ and the corresponding feature vector $\mathrm{AGG}\left(\left\{\phi_{i}\right\}_{i\in\mathcal{S}_{t}}\right)$ is defined to measure the relative improvement to the base feature vector of Eq. (\ref{eq: gap}) in terms of the loss as follows:
\begin{equation}
r_t = \mathcal{L}_t -  \mathcal{L}_\mathrm{GAP}
\label{eq: reward}
\end{equation}
where $\mathcal{L}_t$ is a classification loss of $\mathrm{AGG}\left(\left\{\phi_{i}\right\}_{i\in\mathcal{S}_{t}}\right)$.
With the reward given in Eq. (\ref{eq: reward}), we then define the total return $R_{t}$ as 
\begin{equation}
R_t = \sum_{k=0}^{T'} \gamma^{k} r_{t+k}
\label{eq: return}
\end{equation}
where $\gamma$ denotes a discount factor to deal with a \emph{delayed reward} \cite{sutton2018reinforcement}.

\subsubsection{Actor-Critic Network}
Technically speaking, of various RL approaches, we exploit an actor-critic model \cite{mnih2016asynchronous}, thanks to its popularity and fitness to our problem. That is, our agent maintains a policy network $\pi(a_{t}|s_{t}; \theta_\pi)$ as an actor and a value estimation function $V(s_{t}; \theta_v)$ as a critic. For the $t$\textsuperscript{th} timepoint, the agent receives a state $s_t$ and decides its action $a_t$ from a set of possible actions $\mathcal{A}$ based on the policy $\pi$. Then, the reward $r_t$ and the next state $s_{t+1}$ are obtained from the environment as in Eq. (\ref{eq:action}). 

In our work, we utilize a \emph{synchronized parallel actor-critic} network. Specifically, two distinct deep neural networks are used for a policy estimation and the expected return or  \emph{value} estimation, respectively. The output neurons in our policy network $\pi(a_{t}|s_{t}; \theta_\pi)$ correspond to the probability of taking a selection or rejection action with respect to the current feature vector under the state $s_t$, \ie, $a \sim\pi(a_t|s_t;\theta_\pi)$. Meanwhile, the value estimation network $V(s_{t}; \theta_v)$ has a single output neuron, which produces the expected return under the current state $s_{t}$.



\subsection{Classifier}
After selecting informative feature vectors by our agent over time in a trial, the aggregated vector representation of those is then fed into a densely-connected layer $\rho(\cdot; \theta_\rho)$ for decision-making. 
As for the aggregation, we again introduce the mean average of feature vectors in Eq. (\ref{eq: gap}), also called as the GAP \cite{lin2013network}. In the viewpoint of BCI, the GAP layer can be understood as a means of emphasizing an important spectral range and its neighboring region for each of the feature dimension. 
Using the aggregated feature vector $\mathrm{AGG}\left(\left\{\phi_{i}\right\}_{i\in\mathcal{S}_{T'}}\right)$, the classifier outputs a class label $\hat{\mathbf{y}}$ of the input EEG trial.

\subsection{Optimization and Training Strategy}
To jointly optimize the embedding network, the policy and value networks of an agent module, and a classifier, the proposed framework involves two types of learning schemes, \ie, supervised learning and reinforcement learning. We combine these two learning strategies in our network optimization. 

First, the embedding network $\phi$ and a classifier $\rho$ are pre-trained in a supervised manner without the agent module by minimizing a cross-entropy loss. After pre-training, the actor and critic networks in an agent module are trained to select task-informative features by interacting with the environment. Initially, the agent takes the feature vectors $\boldsymbol{\phi}$ represented by the pre-trained embedding network. Thus, the agent basically starts from the more learned position in a parameter space, rather than a random initial point, thereby training parameters $\phi_{\pi}$ and $\phi_{v}$ faster and more robustly.

The model parameters updating is alternated between (i) the agent module and (ii) the other two modules of feature representation and classification. As the agent is directed to find more informative features by being iteratively updated, the embedding network and the classifier can also focus on the task-oriented feature learning, and thus can be better generalized in a more reliable way. 

To optimize the sequential actions, we update the trainable parameters of the actor network $\theta_\pi$ and the critic network $\theta_v$ by performing a gradient ascent in regard to maximization of the expected total return $\mathbb{E}[R_t]$ ($t=1,...,T'$). 
Basically, the actor parameters $\theta_\pi$ are learned in the direction of $\nabla_{\theta_\pi}\log\pi(a_t|s_t;\theta_\pi)\cdot R_t$ \cite{sutton2018reinforcement}. However, although the updating direction is an unbiased estimate of $\nabla_{\theta_\pi}\mathbb{E}[R_t]$, we need to reduce the variance of this estimate by introducing another value, called \emph{advantage}, \cite{mnih2016asynchronous}. The advantage $A_t$ is calculated as follows:
\begin{equation}
A_t = r_t + \gamma V(s_{t+1};\theta_v) - V(s_t; \theta_v).
\label{eq: advantage}
\end{equation}
By applying the advantage function to the gradient estimation, we define a loss for an actor network as follows:
\begin{equation}
\mathcal{L}_t^\pi = \log\pi(a_t|s_t;\theta_\pi) A_t.
\label{eq: policy_loss}
\end{equation}

Meanwhile, the value estimation function $V(\cdot; \theta_v)$ approximates the expected return for the given state $s_t$, \ie, $V(s_t; \theta_v)=\mathbb{E}[R_t|s_t]$. Owing to the fact that we cannot directly know a value of a specific state, the value estimation function is optimized by a \emph{bootstrapping} method \cite{sutton2018reinforcement}. According to its definition, the current state value estimation $V(s_t; \theta_v)$ should be equal to the summation of the current reward and the next state value estimation $r_t+\gamma V(s_{t+1};\theta_v)$, thus its training loss is defined as follows:
\begin{equation}
\mathcal{L}_t^v = \frac{1}{2}\left[ V(s_t;\theta_v) - \left( r_t + \gamma V(s_{t+1};\theta_v) \right)\right]^2.
\label{eq: value_loss}
\end{equation}

The complete pseudo-algorithm to train all the networks in our framework is presented in Algorithm \ref{alg}.

\begin{algorithm}[t]
	\caption{Pseudo-code for the proposed method \label{alg}}
	\KwIn{Training samples and corresponding labels $\mathbf{x}, \mathbf{y}$}
	\KwIn{Network architectures $\theta_\phi$, $\theta_\pi$, $\theta_v$, and $\theta_\rho$; \# of pre-training $n_\mathrm{pre}$; an optimizer $\mathrm{SGD}$; a learning rate $\alpha$; a discount factor $\gamma$}
	\KwOut{Optimal networks $\theta_\phi^*$, $\theta_\pi^*$, $\theta_v^*$, and $\theta_\rho^*$}
	\For{$i=1,...,n_\mathrm{pre}$}{
		$\boldsymbol{\phi} \leftarrow\phi(\mathbf{x};\theta_\phi)$\;
		$\mathbf{f}_\mathrm{GAP}$$\leftarrow$ Eq. (\ref{eq: gap})\;
		$\hat{\mathbf{y}}\leftarrow\rho(\mathbf{f}_\mathrm{GAP};\theta_\rho)$\;
		Update $\theta_\phi$ and $\theta_\rho$ using $\mathrm{SGD}(\mathrm{BCE}(\mathbf{y}, \hat{\mathbf{y}}), \alpha)$\;
	}
	Estimate $\mathcal{L}_\mathrm{GAP}$ using $\mathbf{f}_\mathrm{GAP}$\;
	\While{Network parameters not converged}{
		$\boldsymbol{\phi}\leftarrow \phi(\mathbf{x};\theta_\phi)$\;
		\For{$t=1,...,T'$}{
			$s_t\leftarrow$ Eq. (\ref{eq: state})\;
			$a_t\sim \pi(a_t|s_t;\theta_\pi)$\;
			$r_t\leftarrow$ Eq. (\ref{eq: reward})\;
			$s_{t+1}\leftarrow$ Eq. (\ref{eq: state})\;
			$\mathcal{L}_t^v\leftarrow$ Eq. (\ref{eq: value_loss})\;
			Update $\theta_v$ using $\mathrm{SGD}(\mathcal{L}_t^v, \alpha)$\;
			$A_t\leftarrow$ Eq. (\ref{eq: advantage})\;
			$\mathcal{L}_t^\pi\leftarrow$ Eq. (\ref{eq: policy_loss})\;
			Update $\theta_\pi$ using $\mathrm{SGD}(-\mathcal{L}_t^\pi, \alpha)$\;
		}
		$\hat{\mathbf{y}}\leftarrow\rho(\mathrm{AGG}\left(\left\{\phi_{i}\right\}_{i\in\mathcal{S}_{T'}}\right);\theta_\rho)$\;
		Update $\theta_\phi$ and $\theta_\rho$ using $\mathrm{SGD}(\mathrm{BCE}(\mathbf{y}, \hat{\mathbf{y}}), \alpha)$\;
	}
\end{algorithm}

\section{Experiments}
\label{sec: experiments}
In this section, we describe the dataset used for performance evaluation, our experimental scenarios, experimental settings, and performance comparison among the competitive methods. 
In regard to the performance comparison, we considered the mean, median and min-max accuracy over all subjects.

\subsection{Dataset and Preprocessing}
We used a publicly available big KU-MI dataset \cite{lee2019eeg}\footnote{Available at \url{http://gigadb.org/dataset/100542}.}, which consists of left-hand and right-hand MI tasks. MI samples were acquired across two sessions from 54 healthy subjects, recorded from 62 Ag/AgCl electrodes according to the standard 10-20 system, and sampled with 1000Hz. Each MI class of the dataset contains 50 trials with a 4-second length. For preprocessing, following \cite{kwon2019subject, lee2019eeg}, we downsampled EEG trials to 100Hz and then applied a band-pass filtering between 8 and 30Hz, including both $\mu$ and $\beta$ bands, and segmented from 1 sec to 3.5 sec (250 timepoints). Finally, we selected 20 electrodes (FC-1/2/3/4/5/6, C-1/2/3/4/5/6/z, and CP-1/2/3/4/5/6/z) over the sensory-motor cortex areas.

\subsection{Experimental Scenarios}
In order to empirically prove the validity of our proposed method, we compare with the existing subject-dependent  and subject-independent methods in performance. By following the recent work of \cite{kwon2019subject}, we set the subject-dependent and subject-independent scenarios as follows:
\subsubsection{Subject-dependent}
For the subject-dependent case, the offline data (training samples) from the second session was used to train the MI classification models. Then, the online data (testing samples) also from the second session was used for the performance validation using the trained models.
\subsubsection{Subject-independent}
For the subject-independent scenario, we conducted a \emph{leave-one-subject-out cross-validation} procedure. To be concrete, we trained subject-independent MI classification models using all training subjects' offline and online data from both sessions. After training, we evaluated the trained models on the target subject's offline data from the second session.

\subsection{Experimental Settings}
\label{subsec: experimental_settings}
While training our proposed framework in Fig. \ref{fig: framework}(a), we set a mini-batch size of 5, an exponentially decreasing learning rate with an initial value of 0.003 and a decreasing ratio of 0.001 per epoch, an RMSProp optimizer \cite{ruder2016overview}, and a Xavier initializer \cite{glorot2010understanding}. For the embedding and classification modules in our framework, we used the existing network architectures of \cite{schirrmeister2017deep, lawhern2018eegnet, ko2020multi}. Briefly, Shallow ConvNet \cite{schirrmeister2017deep} is composed of two convolutional layers, a temporal convolutional layer and a spatial convolutional layer with a square activation function for embedding in a feature space. Deep ConvNet \cite{schirrmeister2017deep} has a temporal convolutional layer, a spatial convolutional layer, and following three temporal convolutional layers with an exponential linear unit (eLU) activation function for feature representation. EEGNet \cite{lawhern2018eegnet} consists of a spectral convolutional layer, a spatial depthwise convolutional layer \cite{chollet2017xception}, and a temporal separable convolutional layer \cite{chollet2017xception} with an eLU activation function for spatio-temporal feature representation. Finally, for the MSNN \cite{ko2020multi}, a spectral convolution and three residually connected temporal separable convolutional layers and spatial convolutional layers with a leaky ReLU function were used as the embedding part. However, in order for better integration with our proposed agent module for signal segments selection, we made a slight modification in the architecture of Shallow ConvNet, Deep ConvNet, and EEGNet by replacing the last feature output layer (\ie, average pooling in Shallow ConvNet and EEGNet, max pooling in Deep ConvNet) with a GAP layer. In this reason, in the following, we differentiate those networks by naming with `original' and `modified' networks. In regard to the classification module, we utilized the above-mentioned networks' densely-connected layers, respectively. As for the SSFR \cite{kwon2019subject}, because it was designed for energy map-based feature representation, rather than the spatio-temporal features, we did not consider it to apply in our framework.


In a pre-training phase for the embedding and classification networks, we set the number of epochs $n_\mathrm{pre}$ by 10.  In regard to the total return $R_{t}$ estimation, a discount factor $\gamma$ of 0.95 was used. For the actor and critic networks, we designed densely-connected layers with a softmax and a sigmoid activation functions for their output layers, respectively. During training, we also applied an elastic net regularizer with the coefficients of  $\ell_1$ and $\ell_2$ as 0.01 and 0.001, respectively. 

We implemented all the models considered in our experiments, except for the linear models and SSFR as their performances were taken from \cite{kwon2019subject}, by Tensorflow 2 \cite{abadi2016tensorflow} and trained on a single Titan RTX GPU on Ubuntu 18.04.

\begin{table}[t]\centering
	\caption{Performance comparison among the comparative and competitive methods under the subject-dependent learning scenario. For the methods with $\star$, their performance was obtained from \cite{kwon2019subject}. AM denotes temporally informative segments selection by our proposed agent module.}
	\begin{tabular}{|c|c|c|c|}\hline
		Method & Mean (SD) & Median & Max-Min\\\hline
		CSP\textsuperscript{$\star$} \cite{blankertz2008optimizing} & 68.57 (17.57) & 64.50 & 100.00-42.00\\
		CSSP\textsuperscript{$\star$} \cite{lemm2005spatio} & 69.68 (18.53) & 63.00 & 100.00-42.00\\
		FBCSP\textsuperscript{$\star$} \cite{ang2008filter} & 70.59 (18.56) & 64.00 & 100.00-45.00\\
		SCCSP \cite{li2017relevant} & 69.13 (16.90) & 64.50 & 100.00-48.00 \\
		BSSFO\textsuperscript{$\star$} \cite{suk2012novel} & 71.02 (18.83) & 63.50 & 100.00-48.00\\
		Shallow ConvNet \cite{schirrmeister2017deep} & 72.39 (16.38) & 68.00 & 100.00-46.00\\
		Deep ConvNet \cite{schirrmeister2017deep} & 62.63 (13.23) & 58.50 & 100.00-50.00 \\
		EEGNet \cite{lawhern2018eegnet} & 64.93 (18.04) & 56.50 & 100.00-47.00\\
		SSFR\textsuperscript{$\star$} \cite{kwon2019subject} & 71.32 (15.88) & 66.45 & 99.00-45.90\\
		MSNN \cite{ko2020multi} & 74.39 (15.59) & 70.50 & 100.00-52.00\\\hline
		Shallow ConvNet + AM & 74.26 (15.76) & 69.00 & 100.00-53.00\\
		Deep ConvNet + AM & 65.02 (15.48) & 58.00 & 100.00-51.00\\
		EEGNet + AM & 67.06 (18.05) & 57.00 & 100.00-50.00 \\
		MSNN + AM &\bf 77.26 (13.92) &\bf 74.50 & 100.00-56.00\\\hline
	\end{tabular}
	\label{table: performance_sbj_dep}
\end{table}

\begin{table}[t]\centering
	\caption{Performance comparison among the comparative and competitive methods under the subject-independent learning scenario. For the methods with $\star$, their performance was obtained from \cite{kwon2019subject}. AM denotes our proposed agent module.}
	\begin{tabular}{|c|c|c|c|}\hline
		Method & Mean (SD) & Median & Max-Min\\\hline
		Pooled CSP\textsuperscript{$\star$} \cite{lotte2009comparison} & 65.65 (16.11) & 58.00 & 100.00-45.00\\
		Fused model\textsuperscript{$\star$} \cite{ray2015subject} & 67.37 (16.01) & 62.50 & 98.00-41.00\\
		MR FBCSP\textsuperscript{$\star$} \cite{lotte2009comparison} & 68.59 (15.28) & 63.00 & 97.00-48.00\\
		SSFR\textsuperscript{$\star$} \cite{kwon2019subject} & 74.15 (15.83) &\bf 75.00 & 100.00-40.00\\
		MSNN \cite{ko2020multi} & 73.96 (17.95) & 73.00 & 100.00-45.00 \\\hline
		MSNN + AM &\bf 75.24 (17.40) &\bf 75.00 & 100.00-45.00\\\hline
	\end{tabular}
	\label{table: performance_sbj_indep}
\end{table}

\subsection{Experimental Results}
\subsubsection{Subject-dependent}
The classification accuracy for the subject-dependent scenario is summarized in TABLE \ref{table: performance_sbj_dep}. First, our proposed method with a modified embedding and classification modules from MSNN \cite{ko2020multi} achieved the highest performance with a large margin, compared to most of the other methods. Second, it is remarkable that deep learning models integrated in our proposed framework for the  embedding and classification modules achieved consistently higher performances than the corresponding original methods. It is also noteworthy that the deep learning models combined with our proposed agent module also enhanced the median and minimum accuracy, compared to their counterparts. This implicitly assures that our proposed framework, especially the agent module, helped to boost the performance across all the subjects.


\subsubsection{Subject-independent}
TABLE \ref{table: performance_sbj_indep} summarizes the classification accuracy of the comparative methods, applicable for the subject-independent scenario. As for our proposed method, we defined the embedding and classification modules with MSNN due to its superiority to other deep models in TABLE  \ref{table: performance_sbj_dep}. Again, our proposed method achieved the highest mean accuracy with a small margin compared to the second best performance by SSFR \cite{kwon2019subject}. It is also noticeable that the use of our proposed agent module helped to enhance the performance by 1.28\% compared to the original MSNN. 


\section{Analyses}
\label{sec: analysis}
In this section, we present the validity of our proposed framework by conducting a statistical test between deep models of involving or non-involving our agent module. We also conduct a qualitative evaluation for the effect of our proposed agent module by comparing (1) the spectrograms of randomly selected EEG signals and our agent-selected EEG signal segments and (2) the topographic maps estimated by full EEG signals and agent-selected signal segments.

\subsection{Statistical Analysis}
\begin{figure}[t]
	\centering{\includegraphics[width=0.9\linewidth]{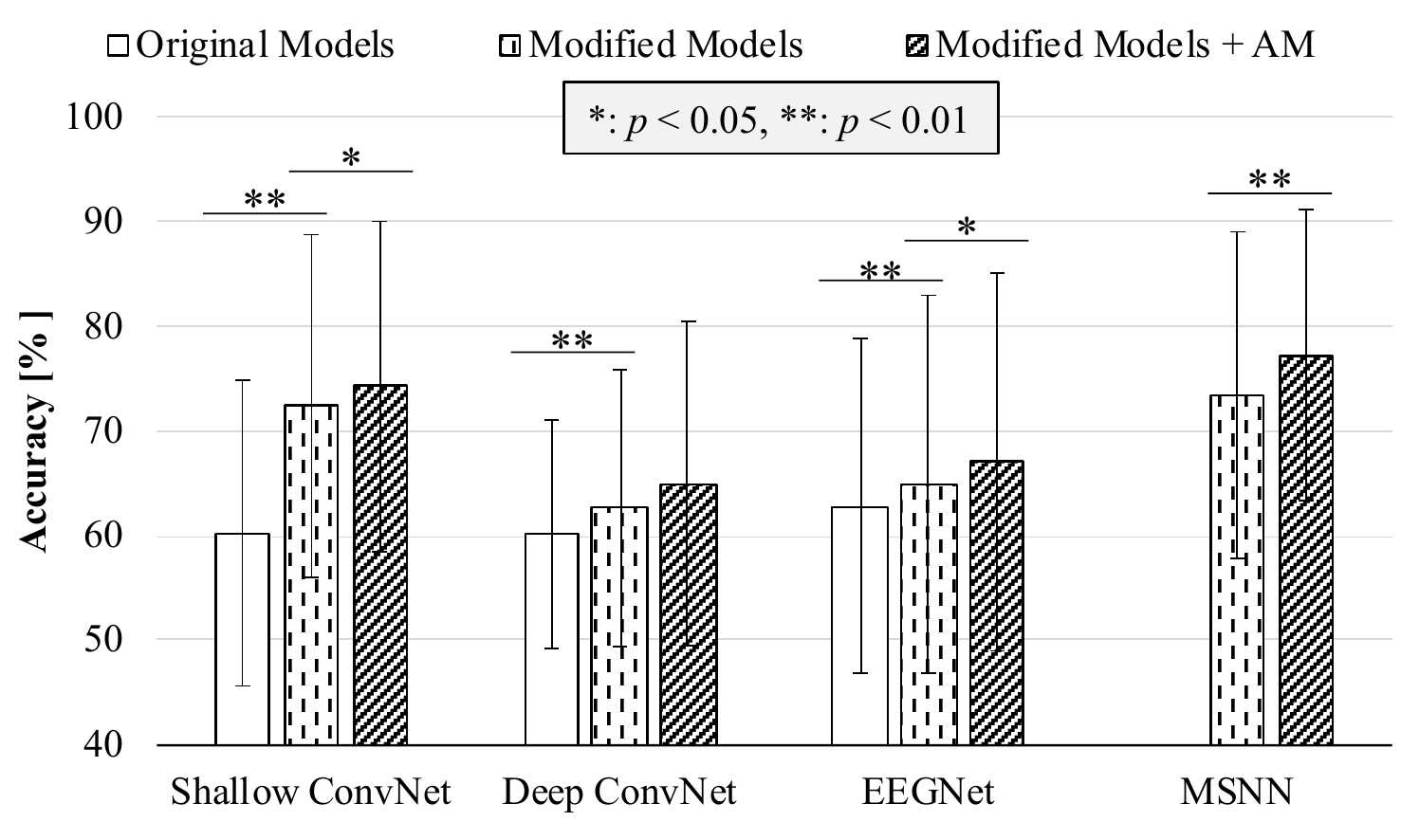}}
	\caption{Performance comparison of MI classification models, `Original Models,' `Modified Models,' and its counterparts trained with the proposed agent module, `Modified Models + AM.'}
	\label{fig: statsitcal_analysis}
\end{figure}

In order to quantitatively validate the effectiveness of our proposed framework, we conducted a two-tailed Wilcoxon's signed-rank test among the original deep models, their modified ones, and the counterpart agent-involved models. The results are plotted in Fig. \ref{fig: statsitcal_analysis}, which state the statistical significance of our proposed agent module with its superiority in classification accuracy. 
In detail, for Shallow ConvNet \cite{schirrmeister2017deep}, EEGNet \cite{lawhern2018eegnet}, and MSNN \cite{ko2020multi}, the proposed framework showed statistical significance with $p$-values of $<0.05$, $<0.05$, and $<0.01$, respectively. From this statistical comparison, it is reasonable to say that our proposed framework, specifically the agent module, played an important role to enhance the classification accuracy across all subjects.  Additionally, we also compared performance of MSNN \cite{ko2020multi} and MSNN combined with our agent module (MSNN+AM) in the subject-independent scenario, and obtained the result that our method was statistically better with $p<0.05$ than the original model in classification accuracy.


\subsection{Qualitative Analysis}
\begin{figure}[t]\centering
	\begin{minipage}[t]{.24\textwidth}
		\begin{center}
			\includegraphics[width=1\linewidth]{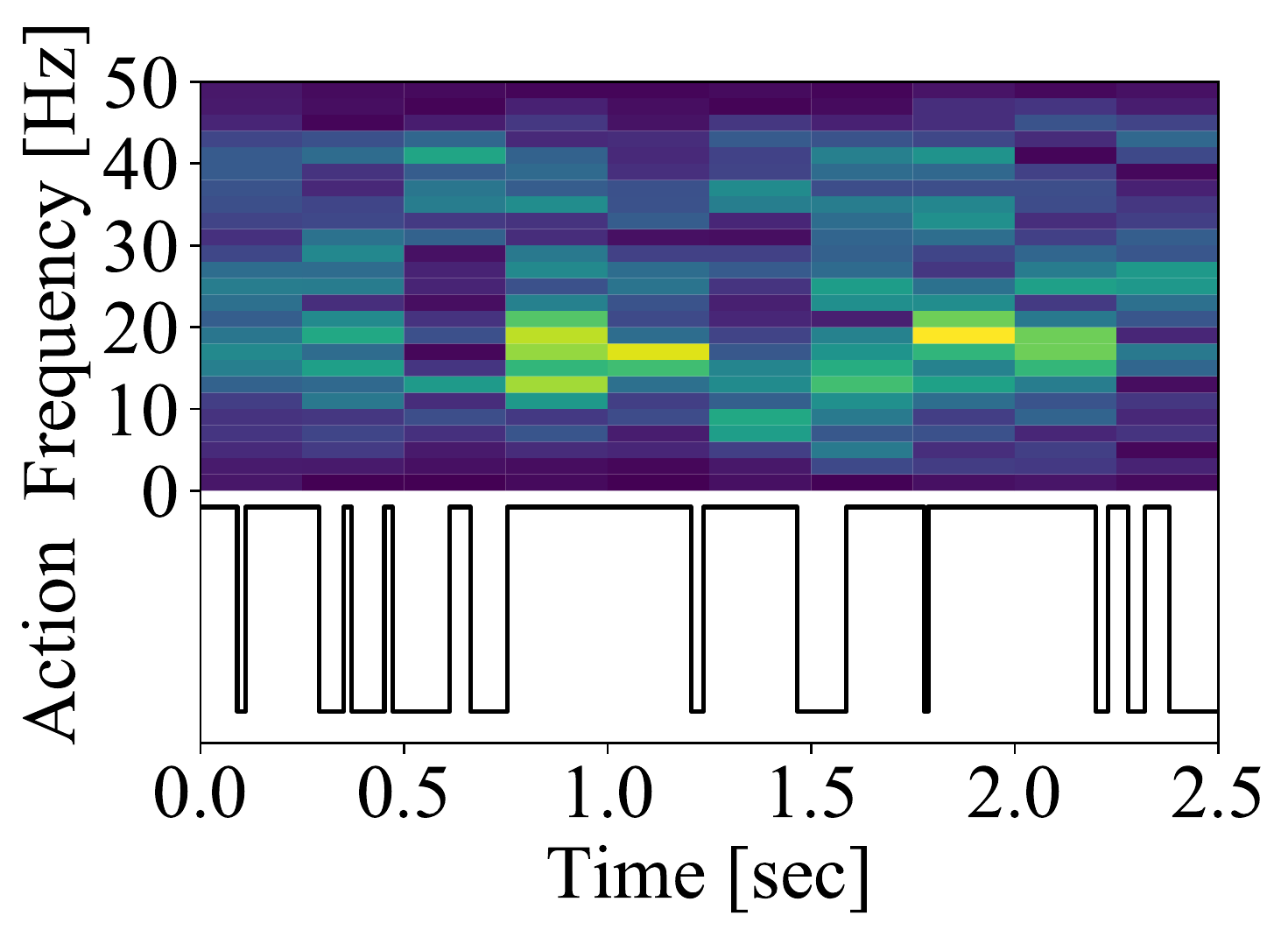}
		\end{center}\vspace{-5pt}
		\footnotesize
		(a) C4 channel signals in a left-hand MI trial from a Subject \#2
	\end{minipage}\hspace{1pt}
	\begin{minipage}[t]{.24\textwidth}
		\begin{center}
			\includegraphics[width=1\linewidth]{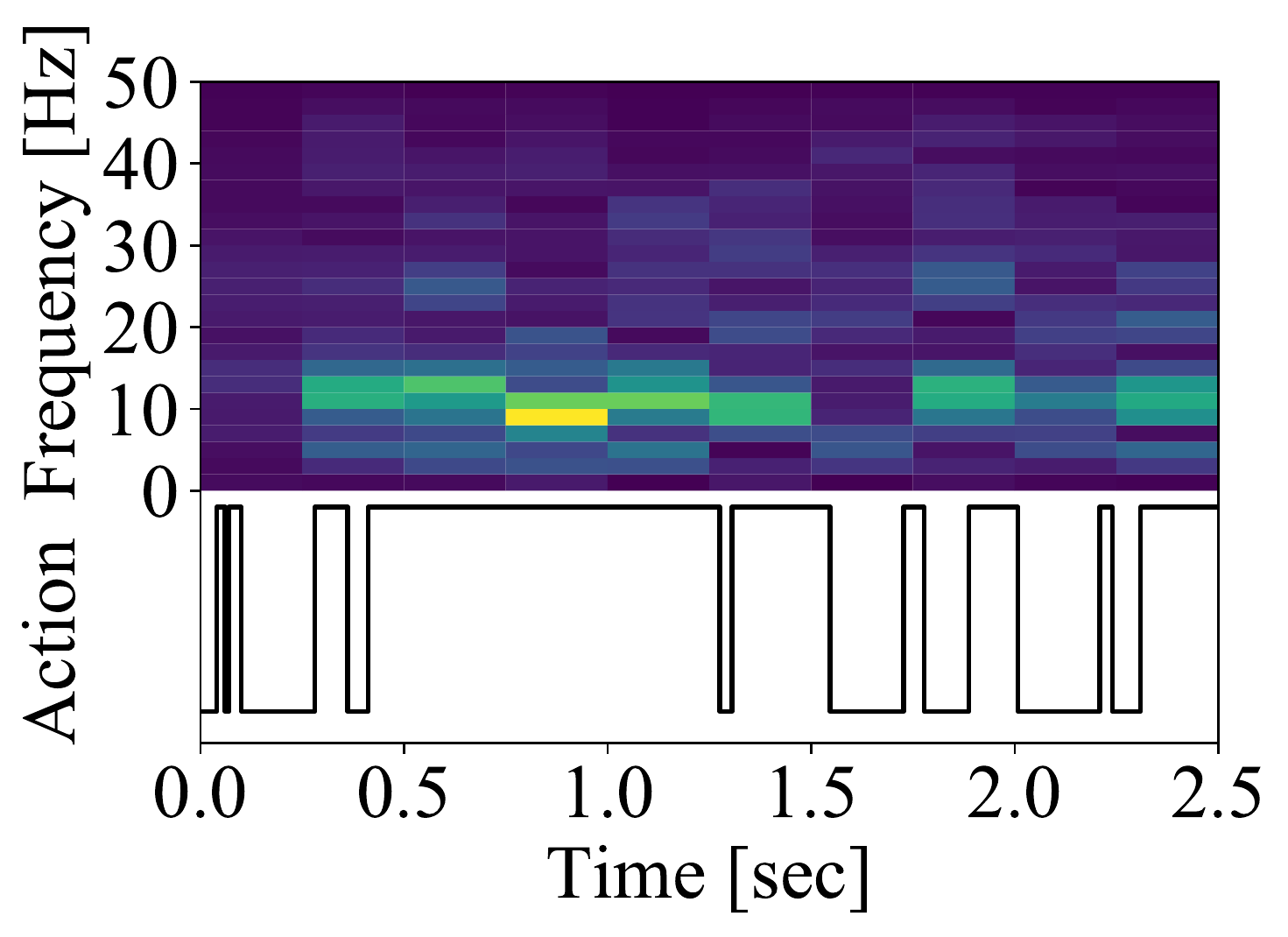}
		\end{center}\vspace{-5pt}
		\footnotesize
		(a) C3 channel signals in a right-hand MI trial from a Subject \#39
	\end{minipage}
	\caption{Spectrograms of the signals measured at C3/C4 channels in the randomly selected trials from two subjects. The bottom graphs represent the a sequence of actions taken by the agent module in our framework with MSNN.}
	\label{fig: selection_analysis}
\end{figure}

In Fig. \ref{fig: selection_analysis}, we visualized the spectrogram (via short-time Fourier transform: STFT) of the C3/C4 channel signals in randomly selected trials from two subjects and the respective action sequences made by our agent module plugged in MSNN. In a coupled-consideration of the power spectrum and the the agent's action of selection, we could observe their positive relations in the sense that the selected signal segments showed high spectral power in the neighbors of the $\mu$ and $\beta$ bands. Note that basically, the timepoints of an agent's view in our framework ($1<t<T'$) is different from the original input timepoints ($1<t<T$) due to a series of convolution operations in the embedding module. For intuitive interpretation of the agent's action, we estimated and aligned the agent's timepoints to the input timepoints by reversely computing the corresponding points in the input space.

\begin{figure}[t]
	\begin{center}
	\includegraphics[width=0.7\linewidth]{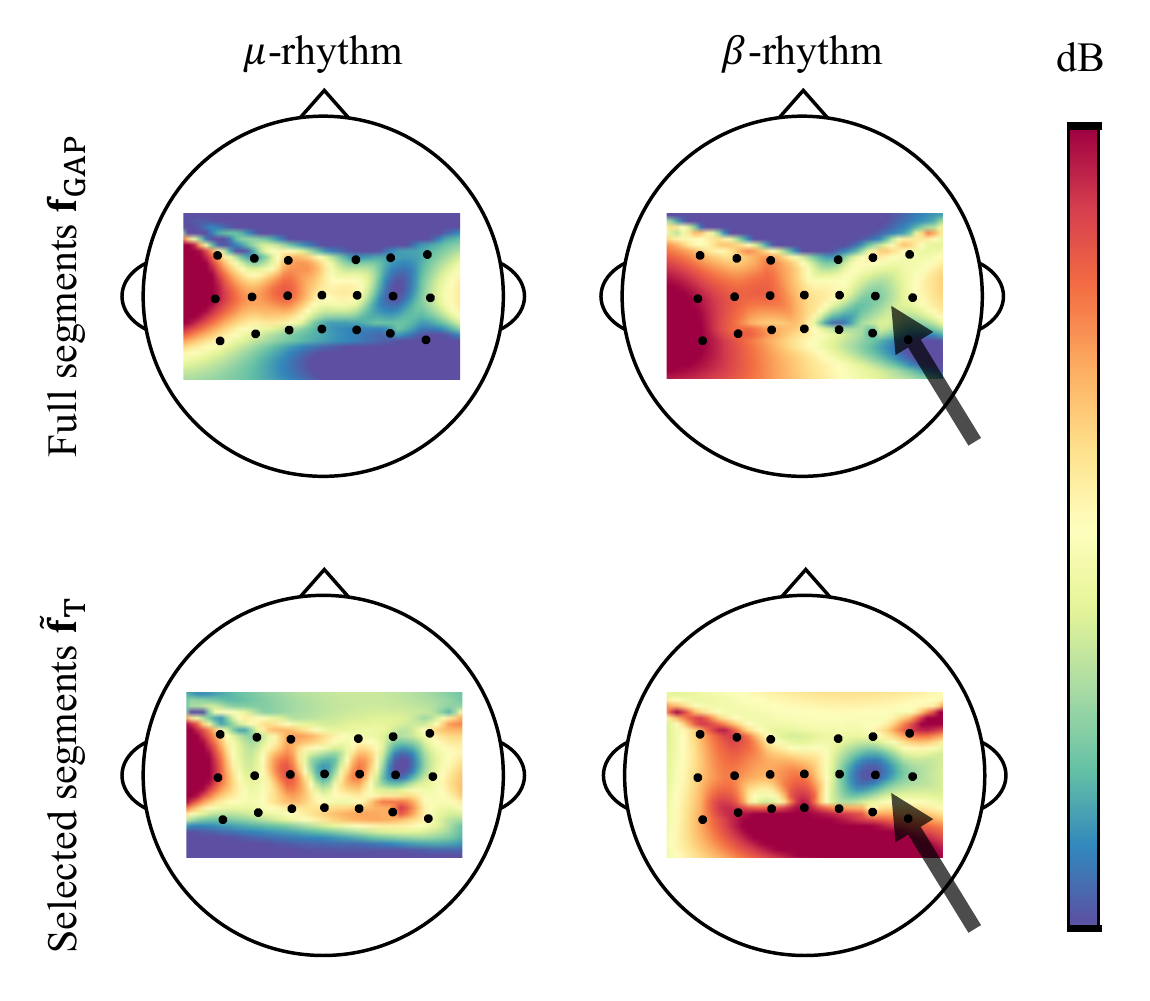}
	\end{center}	
	\footnotesize
	\centering{(a) Topographic maps of a left-hand MI trial from a Subject \#2.}

	\begin{center}
	\includegraphics[width=0.7\linewidth]{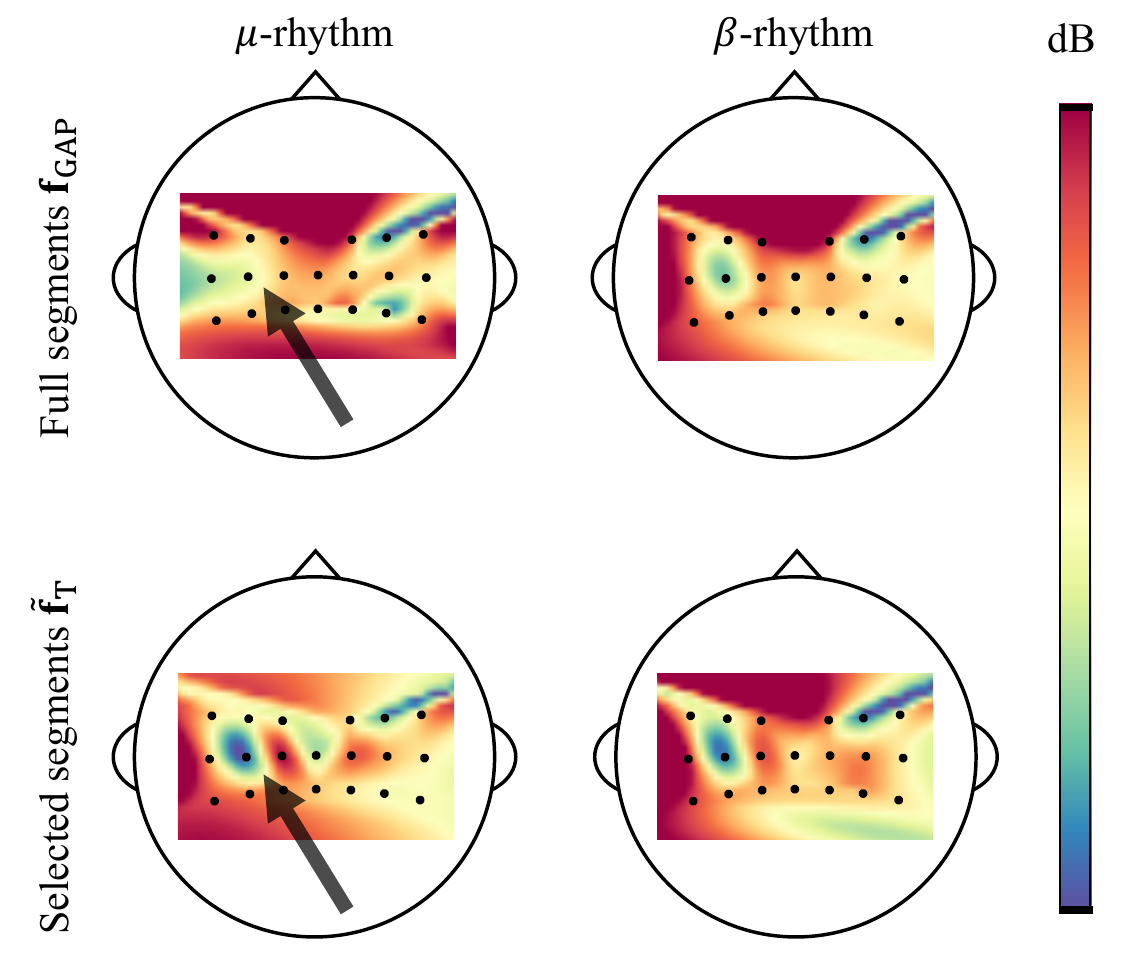}
	\end{center}
	\footnotesize
	\centering{(b) Topographic maps of a right-hand MI trial from a Subject \#39.}
	\caption{Topographic maps of the corresponding spectrograms in Fig. \ref{fig: selection_analysis}. The black dots indicate the montage of the channels and the colors represent the spectral power of the signals observed at the near channels that can be understood as ERD/ERS activations}. 
	\label{fig: topomap_analysis}
\end{figure}

In the meantime, for more neurophysiological inspection, in Fig. \ref{fig: topomap_analysis}, we also visualized topographic maps of full signal segments in a trial and the signal segments selected by our proposed framework for the same trials in Fig. \ref{fig: selection_analysis}. Remarkably, topographic maps based on only the selected signal segments showed more clear and localized ERD/ERS pattern than those from the full signals. In particular, the selected signal segments have more prominent ERD patterns at around the C4 channel in the $\beta$-rhytm than the full signal segments in the Subject \#2. When referring to the spectrogram of that subject in Fig. \ref{fig: selection_analysis}(a), it seemed there was no evident spectral power in the $\beta$-range over the full signal segments in a trial. However, after selecting the task-informative signal segments, we could observe a meaningful and distinguishable local pattern at the C4 channel in the $\beta$-range.
Similarly, in the spectrogram of the full signals in a trial for the Subject \#39 in Fig. \ref{fig: selection_analysis}(b), there seemed less prominent local activations in the $\mu$-range, thus no localized ERD/ERS pattern in Fig. \ref{fig: topomap_analysis}(b). However, after selecting the task-relevant signal segments and plotting the corresponding topographic map, it was then observable a localized ERD/ERS at around the C3 channel.
Based on these results, we empirically conclude that our agent module combined with MSNN in our proposed framework is capable of finding MI-relevant EEG signal segments, thus better learning MI-related feature representations and classifier enhancing the MI classification accuracy. Note that there there was no explicit guide or information for our agent to learn such neurophysiological knowledge.

\section{Conclusion}
\label{sec: conclusion}

In spontaneous BCIs, it is not easy for a user to consistently induce EEG signals for a period of time, apparently for BCI illiterates who are less capable of inducing task-related brain signals. Furthermore, as spontaneous brain signals inducement involves unobservable internal cognitive processes in a brain, it is hard to measure the information level of observed signals with respect to the target tasks, \eg, MI. Hence, it may not for all signals in a trial to necessarily reflect a user's intention.

In this work, we focused on the problem of signals reliability in an MI-EEG trial and proposed a novel framework for task-relevant signal segments selection with an RL-assisted module for better generalization of the trained predictive models. As the components in our proposed framework are modular, it was easy and straightforward to combine with the existing deep models. From our experimental results and analyses over a publicly available big MI dataset, we observed the validity of our proposed method in both quantitative and qualitative comparisons and understandings.

Although we could achieve the state-of-the-art performance in both subject-dependent and subject-independent scenarios in our experiments, there are still some rooms to further improve our method. In particular, the agent module works on a sequence of feature vectors obtained from a preceding embedding module with the full signals in a trial. This mechanism may not be practically useful for online BCIs. Thus, it needs improving the current agent module to be better suited for real-time BCIs, and it will be our forthcoming research issue.



\appendices

\section*{Acknowledgment}

This work was supported by Institute for Information \& Communications Technology Promotion (IITP) grant funded by the Korea government (No. 2017-0-00451, Development of BCI based Brain and Cognitive Computing Technology for Recognizing User's Intentions using Deep Learning). 

This work was also supported by Institute of Information \& communications Technology Planning \& Evaluation (IITP) grant funded by the Korea government (MSIT) (No. 2019-0-00079, Department of Artificial Intelligence (Korea University)).

\ifCLASSOPTIONcaptionsoff
\newpage
\fi

\bibliographystyle{./IEEEtran}
\bibliography{./main}
%
%
%
%
%

\end{document}